# Life-time and line-width of individual quantum dots interfaced with graphene

Xin Miao[1], David J. Gosztola[2], Anirudha V. Sumant[2] and Haim Grebel[1,*]

[1]*Electronic Imaging Center and ECE Dept., New Jersey Institute of technology (NJIT), Newark, NJ 07102, USA. grebel@njit.edu*

[2]*Center for Nanoscale Materials, Nanoscience and Technology Division, Argonne National Laboratory Argonne, IL 60439*

**Abstract:** We report on the luminescence's life-time and line-width from an array of individual quantum dots; these were interfaced with graphene surface guides or dispersed on a metal film. Our results are consistent with screening by charge carriers. Fluorescence quenching is typically mentioned as a sign that chromophores are interfacing a conductive surface; we found that QD interfaced with conductive layers exhibited shorter life-time and line-broadening but not necessarily fluorescence quenching as the latter may be impacted by molecular concentration, reflectivity and conductor imperfections. We also comment on selective life-time measurements, which, we postulate depend on the specifics of the local density-of-states involved.

**I. Introduction:**

Quenching of fluorescence in the vicinity of conductors is well documented [1-2]. The growing interest in graphene [3-5] – a mono, or a few layers of graphite – has extended the study of fluorescence-quenching to this unique film [6-12]. Fluorescence quenching by graphene has been attributed by some to a physical transfer of electrons from the fluorophores to the graphene [6-8], similar to n-type doping in semiconductors. A different point of view was given in [9]; the energy transfer between a Quantum Dot (QD) and graphene was attributed to FRET (fluorescence resonance energy transfer, which could be enabled through screening by free-carriers in the graphene film). These theories do not fully explain the fluorescence quenching in graphene because near the Dirac point such screening is linearly diminishing [10] and the screening, if it exists, should be non-linear and dependent on the amount of charge placed within a small distance away from the graphene [11, 12]. Thus, molecular concentration for each of these independently measured surfaces, as well as the local conductivity of the conductor may be at issue. If fluorescence quenching is due to energy transfer between the chromophore and dipoles in the conductive film, then an increase in the density-of-states for such a radiation outlet is the ultimate proof. Large density-of-states results in shorter life-times and broadening of the fluorescence line [13].

We study isolated QDs: screening by relatively thick QD films and charge coupling between nearby dots may mask the local interaction with the conductor. In order to isolate the QDs from one another we placed each one of them in a hole formed in anodized aluminum oxide (AAO) films. The properties of graphene on periodic and porous substrates, such as AAO have been studied in conjunction with Surface Enhanced Raman (SERS) [14-15] and Surface Plasmon Polariton (SPP) lasers [16-18]. The graphene is partially suspended over the substrate pores. For the energy transfer between the chromophore and graphene to be effective, the characteristic parameter $\alpha=e^2/(\varepsilon\hbar v_F)$ ought to be larger than 1 with $\varepsilon$, the dielectric constant of the vacuum [11]. Also, the absorption of graphene (~2.3% per layer) ought to be compatible to the ~3% absorption of the CdSe/ZnS QD monolayer so that the film of dots will not screen itself [19]. We set to measure life-times and spectral linewidths of QDs interfaced with graphene and with an aluminum film. QDs embedded in a bare AAO hole-array were used as reference.

## II. Experiments

A schematic of our substrate is shown in Fig. 1 along with an SEM picture of QDs embedded in the substrate and coated with graphene, Fig.1b. Details of structure fabrication are provided in the Method section below. The aluminum electrode, was part of the anodization process (Fig. 1c). The QD were coated with a ligand to prevent agglomeration while in suspension; the thickness of the ligand is <8 nm compared to ca 3 nm diameter of the QD.

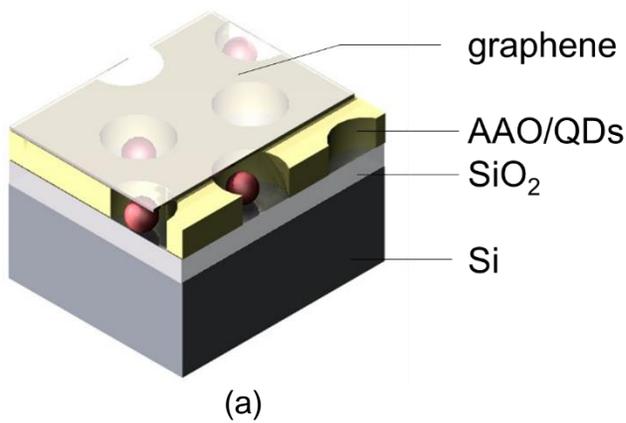

(a)

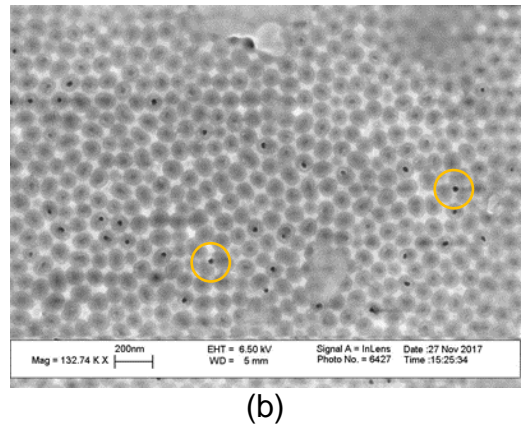

(b)

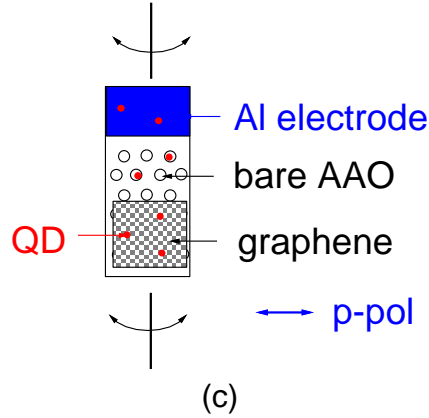

Fig. 1. (a) Schematics of graphene covered configuration. (b) SEM picture of QD-filled hole-array in anodized aluminum oxide (the black dots within the pores. Occasionally one may find QDs, marked by a yellow circle, between the holes. (c) Top view of the sample: the metal electrode, used for anodization, is situated right next to the AAO region. The graphene was covering part of the QD embedded AAO region. The sample was rotated as shown and the incident polarization was p-pol. with respect to sample axis.

**III. Surface modes**

The periodic structure in the AAO regions provides us with an effective way of coupling between surface and radiation modes. This coupling may affect the luminescence intensity as detected by a far-field detector and even its measured life-time constants. Electromagnetic surface modes along the periodic structures may be bound on one side by an effective low index of perforated alumina on the SiO$_2$ layer ($n_{Al2O3/SiO2}$~2) at the sample's bottom. On the other side of the graphene surface guide, the modes may be bound by the low index of either air, or a combination of 200 nm polymer/air layer ($n_{air/polymer}$~1.15); the polymer was a remnant of the graphene transfer process. In calculating the effective indices, we used the relative thicknesses of the various films. An approximation for the refractive index of graphene surface guide may make use of $\varepsilon(\omega)=\varepsilon_b+i\sigma_0/\omega d$: here $\varepsilon_b=5.8\varepsilon_0$ as the effective dielectric constant for graphene with a background material [20] and d=3.38 Angstroms for the effective graphene thickness.

Electromagnetic radiation may be efficiently coupled with a surface mode when the wavevector of either the incident, or scattered (or both) waves are at resonance with the wavevector of the perforated substrate [17]. Since the array pitch is much smaller than the free-space wavelength, a surface mode may become a standing wave, as well. The positions of the QDs are in-phase with the standing electromagnetic surface modes, resulting in enhanced luminescence (Fig. 2).

The tilt angle $\theta$ that produces maximum coupling between the surface and radiation modes may be computed similarly to [21] as,

$$sin(\theta) = \frac{\lambda_0}{a}\sqrt{(\frac{4}{3})(q_1^2 - q_1 q_2 + q_2^2)} - n_{eff} \qquad (1)$$

Here, $\lambda_0$, is the incident or emitted wavelength, $a$, is the pitch for the holes array ($a\sim 90$ nm), $q_1$ and $q_2$ are sub-integers (e.g., 1/3) representing the ratio between the array pitch and the propagating wavelength. Eq. (1) *cannot be fulfilled* for the pump wavelength of 488 nm and $n_{eff}\sim 2.4$ for graphene guide in the range of tilt angles of -8°<$\theta$<8°. Therefore, the fluorescence peaks in Fig. 2 ought to be attributed to the resonances at only *emission* wavelengths. Upon tilting the sample, there are two symmetric peaks in the fluorescence emission as per (1) at ca ±2°.

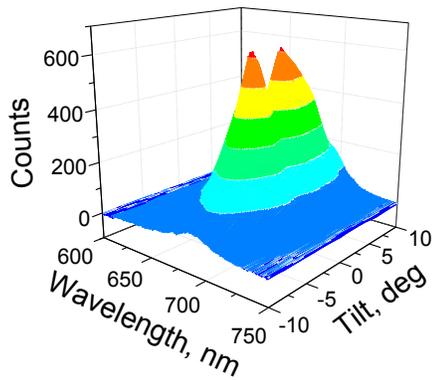
(a)

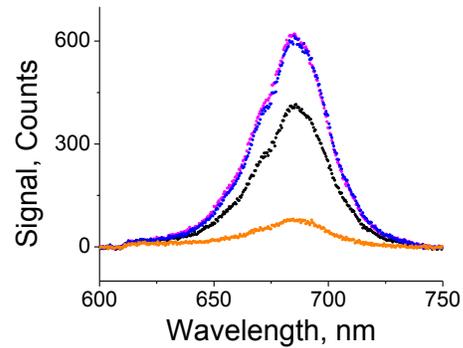
(b)

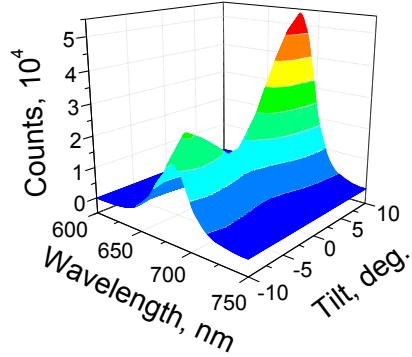

(c)

Fig. 2. (a) Fluorescence of QD690 embedded in graphene covered, AAO hole-array with a pitch of ca 90 nm. (b) A few curves at some specific tilt angles – no meaningful change in the linewidths as a function of tilt angle has been noted. (c) Fluorescence of QD690 embedded in bare AAO.

For QDs embedded in a bare AAO the position of FL peaks has changed to ca ±8º (Fig. 2c). This is consistent with Eq. 1; in absence of graphene, the effective refractive index has reduced, and the angle that satisfies the equation becomes larger.

**IV. Results and Discussions:**

The successful transfer of graphene to the QD loaded AAO hole-array was confirmed by measuring the Raman spectrum of the graphene as shown in Fig. 3. The spectrum was recorded at normal incidence. The relatively small 2D line could be attributed in part to the diminishing quantum efficiency of the Si- based CCD array. In general, the lines have been somewhat blue shifted [22] and could point to the effect of the hole-array.

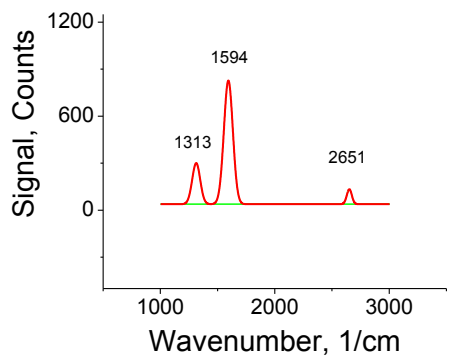

Fig. 3. Raman spectrum of graphene, interfaced with QDs. Data were taken with 11.5 mW 785 nm laser and an x50 LF objective. The small 2D peak is attributed to relatively large defect line at 1313 1/cm (due to contact with the QDs) and the low detector efficiency at that long wavelength (2700 1/cm translates to ~ 950 nm Stokes line).

The photoluminescence life-time measurements were first conducted at normal incidence. The data have been fitted with three time-constants, which fell into three categories: $\tau_1 < 1$ ns; $\tau_2 \sim 1$ ns and $\tau_3 \sim 10$ ns. These correspond to transition rates, b, c, and d respectively (Fig. 4). The longest time-constant (associated with the rate constant d) was still much shorter than those found for QDs in suspension [23]; this may be attributed to the effect of the periodic structure and hence to an increase in the density of states [13].

As noted in [24], the local density of states may be modified by the immediate environment at the chromophore. Thus, our concept of quenching may well be determined by unknown molecular concentration, layer conductivity and the properties of the surface mode. Two examples are shown below: (a) a large transition rates (Fig. 4b) which also portrayed unusual large luminesce; (b) a 'quenched' luminescence (Fig. 5a) which portrayed smaller transition rates (see the caption of Fig. 4).

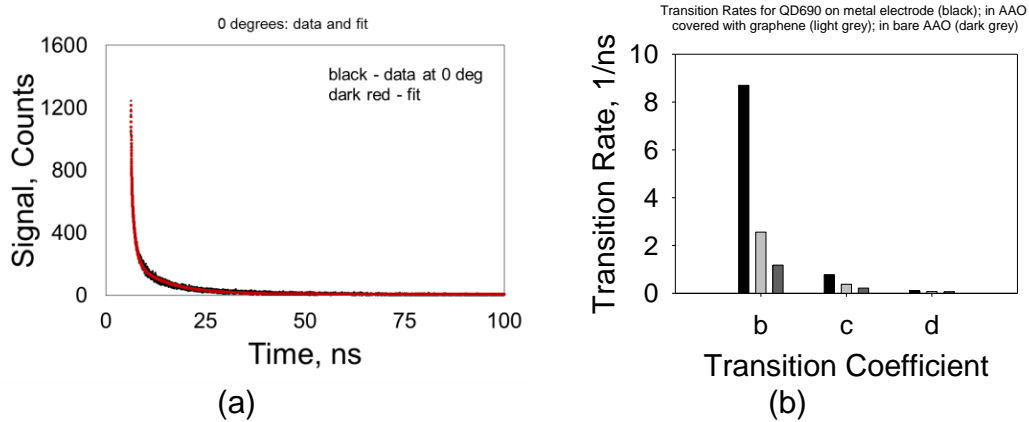

|   | on electrode | in AAO/graphene | in bare AAO |
|---|---|---|---|
| b | 8.7±1.4 | 2.56±0.07 | 1.18±0.04 |
| c | 0.78±0.04 | 0.38±0.012 | 0.22±0.007 |
| d | 0.10±0.001 | 0.07±0.001 | 0.06±0.001 |

Fig. 4. (a) A typical temporal data and its fit at normal incidence (tilt angle, 0°). (b) Various transition rates for QDs: on electrode (black) in AAO hole-array covered with graphene (light grey) and in bare AAO hole-array. The longest life-time was measured for QD embedded in bare AAO where the shortest one was obtained for QD on the aluminum electrode. The table provides with the transition values in 1/ns. The transition values for QDs on the metal are associated with the larger luminescence signal of Fig. 5b. The values for the 'quenched' case (Fig. 5a) are respectively, b=2.58/ns; c=0.37/ns and d=0.07/ns; they are comparable to the graphene values but larger than the values for QDs embedded in bare AAO.

Complementary experiments were conducted on the line broadening of the fluorescence emission (Fig. 5). The spectrum was fitted with two Gaussian peaks whose position and width are provided by the accompanying table. Within the measurement error, no substantial change in the emission linewidth was noted as a function of tilt angle (Fig. 2b). However, as will be seen below, there is a marked change in the related time-constants. One may observe two cases measured for two spots on the electrode: one shown in Fig. 5a is a 'quenched' case, whereas, the one shown in Fig. 5b is an 'enhanced' case. The fluorescence was quenched as expected when the QDs were interfaced with the graphene or the aluminum electrode. This was accompanied by a clear line broadening. We point out that the linewidth of the QDs is masked by an inhomogeneous broadening, attributed to size dispersion.

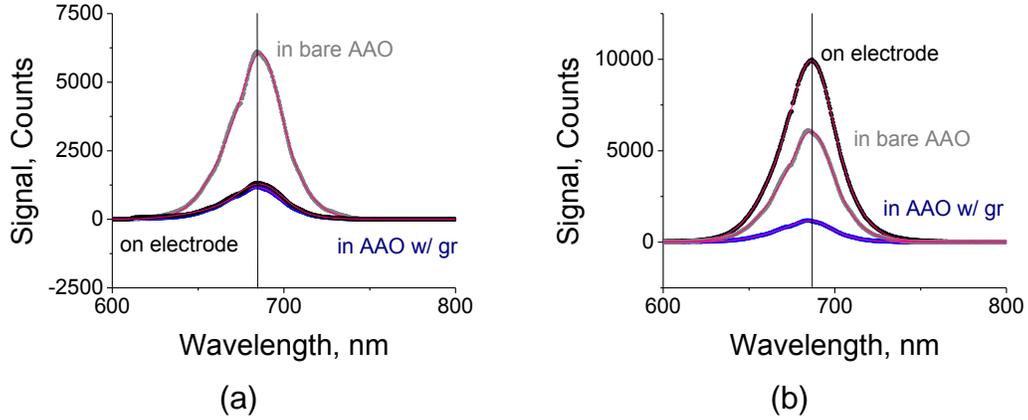

| width (nm) | on Al electrode | peak (nm) | in AAO/graphene | peak (nm) | in bare AAO | peak (nm) |
|---|---|---|---|---|---|---|
| w1 | 50.1±0.51 | 675.36 | 39.63±0.20 | 680.26 | 37.98±0.17 | 682.6 |
| w2 | 25.07±0.25 | 686.72 | 19.85±0.27 | 686.98 | 18.47±0.22 | 687.75 |

Fig. 5. (a) Linewidth of luminescence by QDs on aluminum electrode, in AAO hole-array covered with graphene and in bare AAO hole-array. Quenching of the fluorescence by the graphene and metal is clearly seen. The linewidths for ODs on the electrode or covered with graphene is wider than for QDs imbedded in bare AAO holes. The table summarizes the results. Molecular concentration might be an issue when dealing with luminescence quenching as shown in (b) QDs on a 'hot' metal spot exhibited a much larger signal than the other two cases; nevertheless, the lines widths were respectively, ca 40 nm and 20.6 nm, still larger than the width of QD in bare AAO. The corresponding life-time constants were shorter, as well (Fig. 4, table).

Most puzzling is the increase in the emission photon life-time for QDs interfaced with graphene at tilt angles that seem to be associated with resonance coupling between the surface and the emission modes. In Fig. 6 we show the various rate coefficients as a function of tilt angle. One expects that when at resonance, the measured emission would exhibit a shorter life-time due to an increase in the density of states of its surface modes [13]. Similar experiments with QDs in bare AAO yielded much smaller luminescence changes (less than 3% in the transition coefficients compared with a larger than 10% change for luminescence of QDs interfaced with graphene coated AAO) and therefore deemed inconclusive. Nevertheless, coupling to the radiation modes is strong as observed in Fig. 2c.

Fermi's Golden rule relates the transition rate of the QDs to the final density of states at the emission frequency. In principle, the emission from a QD may be funneled through several radiation venues (waveguide modes, resonance modes, surface modes, etc.,) each of which has a different local, or global density-of-states (DOS). These venues are not necessarily coupled together and the impact of their density-of-states may not be simply summed up as was done in [24]; the photon has a finite probability to decay via each of these outlets. In the case of graphene surface guides, tilting of the sample resulted in capturing a subset of these venues, e.g., decay through a collective surface guiding mode, whose density of states is smaller than the one that was measured at off-resonance [25]. Specifically, the DOS for a two-dimensional propagating surface guide is linearly proportional to the radial frequency, ω, whereas the DOS for a three-dimensional free space radiation mode is proportional to the radial frequency squared, $ω^2$. Thus, in principle, at off-resonance conditions, the emission from a single QD emitter may couple to a larger density of states pool, and therefore exhibits a shorter life-time. As stated before, inhomogeneous line broadening as a result of QD size dispersion may have obscured linewidth effects as a function of tilt. All of that means that graphene is better at sustaining surface propagating modes.

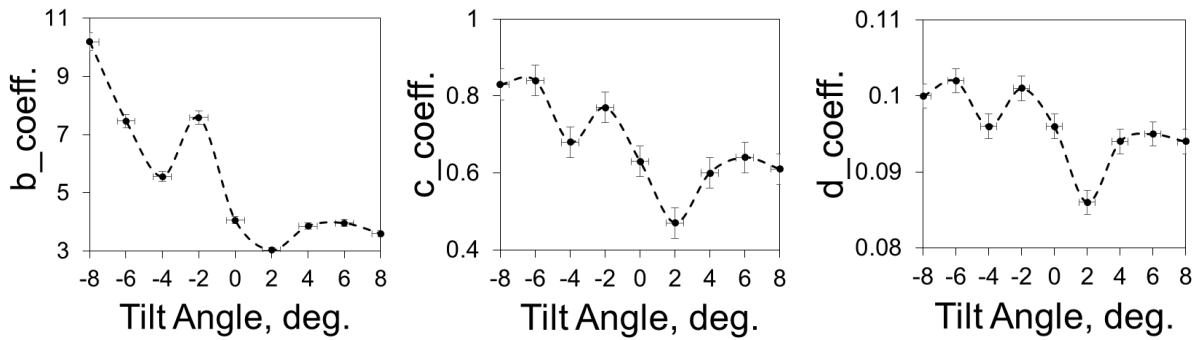

Fig. 6. The rate coefficients as a function of tilt angle. Close to resonance coupling, these coefficients are at the minimum (suggesting longer photon life-time). While there are variations due to local imperfections, the trend, as judged by the coefficients on either side of the minimum is nonetheless clear. The connecting dash curves are only guide to the eye.

In summary, we measured life-time and linewidth for QDs on aluminum electrode, in AAO hole-array interfaced with graphene and compared it with QDs embedded in bare AAO hole-array. Indeed, QD interfaced with conductive films portrayed shorter life-time and line-broadening but not necessarily fluorescence quenching.

**Methods:** 20 nm of $SiO_2$ of thermal oxide was deposited on a <100> p-type 1-10 Ohms.cm Si wafer. For the anodization, a 1-μm Al film was deposited on top of the $SiO_2$ layer; the Al was later anodized completely per previous recipe [16] – its final thickness was estimated as ~50 nm. Anodization of the Al resulted in a hole-array with a pitch of ca 90 nm and a hole-diameter of ca 20 nm. The hexagonal hole-array was polycrystalline with a typical domain size of ~10 μm. The CdSe/ZnS QDs with peak luminescence of ca 690 nm were suspended in toluene and drop-cast into the anodized porous substrate. The QDs were coated with octadecylamine to prevent agglomeration while in suspension. Mostly one QD occupied an AAO nano-hole (Fig. 1b) with an estimated concentration of 25 QDs/ $μm^2$. Excess dots lying on the substrate surface were washed away by rinsing with toluene.

The graphene was produced by chemical vapor deposition (CVD) on copper foil and was transferred onto the QD embedded substrate by use of 200 nm poly(methyl methacrylate), PMMA film [26]. In some cases we retained the PMMA film as a protective upper coating. The presence of the PMMA did not affect the life-time nor the spectral line widths.

Life-time and spectral line width data were obtained using a microscope system (Olympus IX71) coupled to both a spectrometer with a CCD detector array and to a single photon avalanche photodiode (SPAD). The sample was excited with 488 nm pulses (19 µW, 5 MHz, 200 ps) from a supercontinuum laser (Fianium WhiteLase SC-390). The excitation wavelength was selected using an acousto-optic tunable filter (AOTF) along with a bandpass filter. A 5x objective (Olympus NeoSPlan, 0.13 NA) was used to both focus the excitation and collect the emission. A dichroic filter (Semrock FF506-Di03) was used to separate the excitation and emission wavelengths. For spectral measurements, the collected emission was directed to the entrance slit of a 300 mm focal length spectrometer (Acton, SP2300) equipped with a 150 l/mm diffraction grating and a 1320 x 100 channel CCD (Princeton Instruments, PIXIS 100BR). Time-resolved data was collected using the time-correlated single photon counting technique (TCSPC). For the TCSPC measurements, the collected emission from the sample was sent to a SPAD (MPD SPD) after passing through a long pass filter (Chroma, HQ520LP). The pulses from the SPAD were recorded using a computer controlled TCSPC system (Picoquant, PicoHarp300). For the angle-resolved measurements, the sample was tilted with respect to the p-polarized laser (Fig. 1c).

Tilting of the sample was made by modifying the optical microscope to include a rotational stage instead of the tradition microscope platform. The spot position of the focused 488 nm pump beam was monitored by a separate CCD camera to help minimizing spot wobbling. Due to the relatively large pump spot, re-focusing was found un-necessary for angles smaller than 10 degrees; however, this may be of concern for tightly focused beams.

**Acknowledgement:** This work was performed, in part, at the Center for Nanoscale Materials, a U.S. Department of Energy Office of Science User Facility, and supported by the U.S. Department of Energy, Office of Science, under Contract No. DE-AC02-06CH11357.